\documentclass[journal]{IEEEtran}

\usepackage{graphicx}
\usepackage{cite}
\usepackage{units}
\usepackage{amsmath}
\usepackage{color}

\usepackage{lipsum}
\usepackage{hyperref}

\usepackage[pscoord]{eso-pic}

\newcommand{\placetextbox}[3]{
  \setbox0=\hbox{#3}
  \AddToShipoutPictureFG*{
    \put(\LenToUnit{#1\paperwidth},\LenToUnit{#2\paperheight}){\vtop{{\null}\makebox[0pt][c]{#3}}}%
  }%
}%

\DeclareMathOperator*{\E}{\large \textrm{E}}

\newcommand{\xv}{\mbox{\boldmath$x$}}
\newcommand{\zv}{\mbox{\boldmath$z$}}

\newcommand{\yv}{\mbox{\boldmath$y$}}
\newcommand{\Hv}{\mbox{\boldmath$H$}}
\newcommand{\Av}{\mbox{\boldmath$A$}}

\newcommand{\etav}{\mbox{\boldmath$\eta$}}
\newcommand{\xiv}{\mbox{\boldmath$\xi$}}

\newcommand{\fv}{\mbox{\boldmath$f$}}
\newcommand{\gv}{\mbox{\boldmath$g$}}
\newcommand{\nuv}{\mbox{\boldmath$\nu$}}
\newcommand{\Iv}{\mbox{\boldmath$I$}}
\newcommand{\hv}{\mbox{\boldmath$h$}}
\newcommand{\muv}{\mbox{\boldmath$\mu$}}
\newcommand{\Sigmav}{\mbox{\boldmath$\Sigma$}}

\newcommand{\wv}{\mbox{\boldmath$w$}}
\newcommand{\varrhov}{\mbox{\boldmath$\varrho$}}

\newtheorem{remark}{Remark}

\title{Approximating Optimal Estimation of Time Offset Synchronization with Temperature Variations
\thanks{This work was partially supported
    by the Ministry of Education, University, and Research of Italy (MIUR)
		in the framework of the
    Project Wi-Fact ``WIreless FACTory and beyond'' (DM53543).}}

\author{Maurizio~Mongelli,~\IEEEmembership{Member,~IEEE}, Stefano~Scanzio

}

\begin{document}
\placetextbox{0.5}{1}{This is the author's version of an article that has been published in this journal.}
\placetextbox{0.5}{0.985}{Changes were made to this version by the publisher prior to publication.}
\placetextbox{0.5}{0.97}{The final version of record is available at \href{https://doi.org/10.1109/TIM.2014.2320400}{https://doi.org/10.1109/TIM.2014.2320400}}%
\placetextbox{0.5}{0.05}{Copyright (c) 2014 IEEE. Personal use is permitted.}
\placetextbox{0.5}{0.035}{For any other purposes, permission must be obtained from the IEEE by emailing pubs-permissions@ieee.org.}%

\maketitle

\begin{abstract}
The paper addresses the problem of time offset synchronization in the presence of temperature variations, which lead to a non-Gaussian environment. In this context, regular Kalman filtering reveals to be suboptimal. A functional optimization approach is developed in order to approximate optimal estimation of the clock offset between master and slave. A numerical approximation is provided to this aim, based on regular neural network training. Other heuristics are provided as well, based on spline regression. An extensive performance evaluation highlights the benefits of the proposed techniques, which can be easily generalized to several clock synchronization protocols and operating environments.
\end{abstract}

\begin{IEEEkeywords}
Synchronization,
Clocks,
Control systems,
Neural networks,
Real time systems
\end{IEEEkeywords}

\section{Introduction}
\IEEEPARstart{C}{l}ock Synchronization Protocols (CSPs) are becoming a fundamental component of industrial networks as they are widely used in all the contexts in which a common time reference is required \cite{2013-IEM-CSP1, 2013-IEM-CSP2}. The notion of time usually follows a clock register, which is cyclically updated with a frequency derived from a Crystal Oscillator (XO). Clock registers may differ in virtue of changes in XOs oscillation frequencies due to time-varying environmental conditions involving temperature, power supply, vibrations, humidity or pressure \cite{2004-TUTORIAL-VIG}. The purpose of a CSP is exchanging synchronization messages to keep the clock registers of different nodes aligned, by using timestamps on such messages. The synchronization error thus depends also on the rate at which the synchronization messages are exchanged and on their noisy content (timestamps \emph{jitter}) due to unreliability of both communication and messages processing.
\subsection{Motivation}
Synchronization algorithms are used to reduce the synchronization error by exploiting the information contained in the timestamps. At this purpose, several CSPs are based on Kalman filtering, in particular with respect to the IEEE 1588 protocol \cite{kalman_bello2, ToIM, Siemens, kalman_bello}, which is the ``de facto'' standard in this context. Similar Kalman-based approaches have been also investigated for other CSPs, such as the Network Time Protocol (NTP) \cite{KalmanNTP}.
The fundamental hypotheses of Kalman estimation are: linear dynamics of the system, Gaussian noise components for both state and measurements variables and a-priori knowledge of the noise covariances.
\subsection{Contribution}
XOs are usually modeled by linear equations \cite{kalman_bello2, ToIM}, but non-Gaussian noises lead to sub-optimal performance as stated in \cite{kalman_bello2}. In this perspective, this paper formulates an optimal estimation approach beyond regular LQG hypotheses \cite{Siemens} (i.e., linear dynamics, quadratic cost function, Gaussian noises). The key idea is deriving the functional optimization formulation of optimal estimation and computing the inherent solution via neural approximation \cite{Zoppoli}. The main focus relies on the frequency variation component of the clock, whose noise is affected by temperature changes and other effects outside the control of the system designer, which are far from being Gaussian. Despite the approach presented in this paper does not guarantee optimal estimation, it outperforms Kalman filtering and infers the right synchronization correction over a large set of experimental conditions, without any ad-hoc adjustment of the algorithm' parameters. The approach can be generalized to other non-LQG conditions, as the one of \cite{RTSPWSNs}, in which non-linear clock models are claimed to be more accurate for Wireless Sensor Networks (WSNs). It must be noted, however, that the WSNs context deserves special attention also with respect to the energy efficiency of the algorithm, which this paper does not address explicitly.

The following section deals with the used CSP model, which is updated by an additional component concerning the XO exposed to temperature variations. In Section \ref{sec:optimalFilter}, the concept of optimal filter is formulated. Kalman filtering is addressed by Section \ref{sec:kalman}. The proposed approaches are: high-order regression splines and neural approximation of the optimal filter. They are addressed by Sections \ref{sec:splines} and \ref{sec:neuralApproximation}, respectively. The experimental setting is discussed in Section \ref{sec:experimentsSetting}. Simulation results are provided in Section \ref{sec:performanceEvaluation}. Section \ref{sec:conclusions} concludes the paper and summarizes possible topics of future research.

\section{Clock model and correction}
The clock model considered here slightly differs from the one of \cite{ToIM}. At a generic time $k$, the clock registers of the master and the slave nodes (denoted by $C_M(k)$ and $C_S(k)$, respectively) differ of the offset quantity $\theta(k)$.
\begin{eqnarray}
C_S(k) = C_M(k)+\theta(k) \label{eq:CSCM}
\end{eqnarray}
A typical two states clock model can be represented by the following equations in the discrete domain:
\begin{eqnarray}
\theta(k) & = & \theta(k-1) + \gamma(k-1) \cdot \tau + \omega_{\theta}(k-1) \label{eq:state1}\\
\gamma(k) & = & \gamma(k-1) + \omega_{\gamma}(k-1) + \omega_{\gamma^{T}}(k-1,\cdot) \label{eq:state2}
\end{eqnarray}
where $\tau$ is the size of the time step and $\gamma(k)$ represents the \textit{skew} variations, as commonly represented in the scientific literature. The \textit{skew} defines the normalized difference between the XO oscillation frequency and its nominal frequency. The $\omega_{\theta}$ and $\omega_{\gamma}$ quantities represent the noises affecting $\theta$ and $\gamma$, respectively, whose distributions are of the Gaussian type; the inherent standard deviations are denoted by $\sigma_{\omega_{\theta}}$ and $\sigma_{\omega_{\gamma}}$, respectively. The $\omega_{\gamma^{T}}$ quantity represents the noise effect due to temperature variations as detailed later.

CSPs without the compensation of the propagation delay, such as the \textit{sender-receiver} FTSP \cite{2004-FTSP} or the \textit{receiver-receiver} RBS \cite{2002-RBS} and RBIS \cite{2012-ETFA-RBIS}, estimate the offset $\theta(k)$ and the skew $\gamma(k)$ using two timestamps. Without any synchronization assistance, the simplest way to update the two quantities at the slave is:
\begin{eqnarray}
\hat\theta(k) & = & \hat T_S(k)-\hat T_M(k) \label{eq:measure1}\\
\hat\gamma(k) & = & \frac{\hat\theta(k) - \hat\theta(k-1)}{\hat T_M(k)-\hat T_M(k-1)} \label{eq:measure2}
\end{eqnarray}
having the timestamps $\hat T_M(k)$ and $\hat T_S(k)$ acquired on a common time event $k$ at master and slave. The timestamps are affected by jitters $\omega_M$ and $\omega_S$, whose standard deviations are denoted by $\sigma_M$ and $\sigma_S$, respectively:
\begin{eqnarray} \label{eq:measurementnoises}
\hat T_M(k) & = & C_M(k) + \omega_M(k) \label{eq:measurementnoises1} \\
\hat T_S(k) & = & C_S(k) + \omega_S(k) \label{eq:measurementnoises2}
\end{eqnarray}
The analysis made in \cite{SP} highlights the separate effects of both the frequency skew and the jitter. CSPs with the compensation of the propagation delay (such as the IEEE 1588 \cite{2008-IEEE-1588std} and RTSP \cite{RTSPWSNs}) make use of 4 timestamps to estimate the propagation delay. In such a case, equations (\ref{eq:measure1}) and (\ref{eq:measure2}) are slightly different, see, e.g., \cite{ToIM}. The algorithms formulated in this work take (\ref{eq:measure1}) and (\ref{eq:measure2}) as a reference and are applicable to other CSPs (i.e., sender-receiver, receiver-receiver, with and without the estimation of the propagation delay), by updating (\ref{eq:measure1}) and (\ref{eq:measure2}) in agreement with the specific characteristics of the CSP \footnote{The three state equation model of \cite{kalman_bello2} can be incorporated in the following formulations as well, without significant changes to the derivation of the proposed algorithms.}.

The distribution of jitters is supposed here Gaussian to focus on non-Gaussian temperature variations, but the hypothesis may be questionable, especially with software timestamping.
\subsection{Temperature Variations} \label{sec:TemperatureVariations}
\begin{figure*}
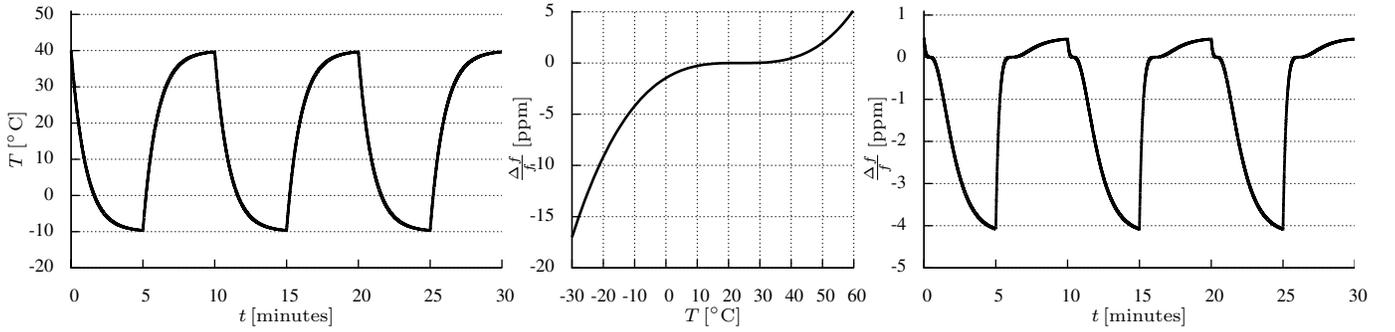

  \scriptsize
  \centering
  \include{fig1}
  \caption{Time-temperature variation, temperature-frequency characteristic, and time-frequency variation plots with $p=\unit[600]{s}$ and $tc=\unit[60]{s}$.}
  \label{fig:graphTempVar}
\end{figure*}
The model of the temperature variations takes as a reference a scenario in which an object changes different environments, characterized by different temperatures. A typical example deals with a Programmable Logic Controller (PLC) in a mobile carriage that moves inside and outside a warehouse. Other examples are embedded systems placed near a heat source, such as a car engine, or exposed to a transition between a state of sunshine and a state of shadow exposition. In these and other typical situations, the temperature variations can be fast, thus causing wide variations of the XO frequency. \\ Let two consecutive temperature changes have a round trip period $p$. When the object is placed in an environment with a different temperature, the temperature $T$ of the XO varies with the following law\footnote{The temperature variation of an object moved to an environment with a different temperature can be modelled by the so-called \textit{Newton's law of cooling}. If the object has a negligible internal thermal resistance, the law is a precise approximation of its temperature evolution over time. The hypothesis of negligible thermal resistance may be reasonable for a XO.}:
\begin{eqnarray}
T = T_E+(T_{XO}-T_E) \cdot e^{-\frac{1}{tc}\Delta t}
\end{eqnarray}
where $T_{XO}$ is the initial temperature, $tc$ is the \textit{thermal time constant} (i.e., the time needed to reach the $\unit[63.2]{\%}$ of $T_E$, in relation with the initial temperature $T_{XO}$) and $T_E$ is the temperature of the environment in which the XO is moved on. The left plot of Fig. \ref{fig:graphTempVar} represents the temperature variation of a XO contained in a shell with $tc=\unit[60]{s}$; it is periodically moved every $p=\unit[600]{s}$ between two environments with temperatures $T^{High}_E=\unit[40]{^{\circ}C}$ and $T^{Low}_E=\unit[-10]{^{\circ}C}$, respectively. Different types of XOs with different characteristics are available. Almost the totality of off-the-shelf PCs and embedded systems make use of AT-cut XOs to beat time, since they are very cheap. Unfortunately, such a kind of XOs highly suffer of oscillation frequency variations with respect to temperature. Temperature compensated XOs exist (VCXO, OCXO, TCXO, TVVCXO, OCVCXO), but usually they are not installed in common systems. For this reason, temperature variation is one of the most important causes of frequency instability on today devices \cite{2004-TUTORIAL-VIG}.
The temperature-frequency relation for AT-cut XOs \cite{1962-IRE-XO} can be accurately expressed with a $3$rd polynomial form:
\begin{eqnarray}
\frac{\Delta f}{f} = a\cdot(T-T_0) + b\cdot(T-T_0)^2 + c\cdot(T-T_0)^3 \label{eq:tempFreqChar}
\end{eqnarray}
where $T_0$ is the reference temperature of the XO and $a$, $b$ and $c$ three coefficients that represent the XO characteristic. The values used in this work are $T_0=\unit[25]{^{\circ}C}$, $a=0.0$, $b=0.4 \cdot 10^{-9}$ and $c=109.5 \cdot 10^{-12}$, as reported in \cite{1962-IRE-XO}, and the temperature-frequency characteristic is reported in the middle plot of Fig. \ref{fig:graphTempVar}. From the time-temperature plot (left in Fig. \ref{fig:graphTempVar}) and the temperature-frequency function (\ref{eq:tempFreqChar}), the noise component due to temperature $\omega_{\gamma^{T}}(t,\cdot)=\omega_{\gamma^{T}}(t,tc,p,T^{High}_E,T^{Low}_E)$ can be eventually derived and quantized. The inherent probability distribution is not Gaussian; it is actually a multi-modal distribution, with significant asymmetry among the peaks.

\section{Optimal Filter}\label{sec:optimalFilter}
Let $\xv(k)=\fv(\xv(k-1),\xiv(k-1))$ be the state equation in compact form from (\ref{eq:state1}) and (\ref{eq:state2}) with $\xv(k)=[\theta(k), \gamma(k)]$, $\xiv(k)=[\omega_\theta(k), \omega_\gamma(k)+\omega_{\gamma^{T}}(k,\cdot)]$ and $\yv(k)=\gv(\xv(k),\etav(k))$ be the measurement equation from (\ref{eq:measure1}) and (\ref{eq:measure2}) with $\yv(k)=[\hat \theta(k), \hat \gamma(k)]$, $\etav$ being the vector of measurement noises in (\ref{eq:measurementnoises1})-(\ref{eq:measurementnoises2}) on $\theta$ and $\gamma$, respectively. Both $\fv(\cdot)$ and $\gv(\cdot)$ are linear functions and can be expressed in terms of basic matrix algebra \footnote{$\xv(k)=\Av \xv(k-1)+\xiv(k-1)$, $\Av=[1, \tau; 0, 1]$; $\yv(k)=\Hv \xv(k)+\etav(k)$, $\Hv$ being the identity matrix.}. The optimal estimator (or optimal filter) is defined by the law $\nuv_k^o(\cdot)=\nuv_k^o(\Iv_k)$ minimizing the following functional cost:
\begin{eqnarray} \label{eq:OptimalFilter}
\nuv_k^o(\Iv_k) = arg \min_ {\nuv_k(\Iv_k)} \E_{\scriptsize {\xv(k)}} \{ \hv(\xv(k)-\nuv_k(\Iv_k))  | \Iv_k \}, \forall \ \Iv_k
\end{eqnarray}
$\Iv_k$ being the information vector collecting all the measurements from the beginning $\Iv_k=[\yv(0),...,\yv(k)]$ and $\hv(\cdot)$ being a Bayesian risk function \footnote{$\hv(\zv)$ is a Bayesian risk function if the following are met: $\hv(\zv)$ is not negative, it is symmetric, i.e., $\hv(\zv)=\hv(-\zv)$ and it is not decreasing with increasing positive \zv; in the scalar case, examples are: $h(z)=z^2$ and $h(z)=|z|$.}. The optimal filter thus implies the solution of a functional optimization problem. Formally, a new instance of problem (\ref{eq:OptimalFilter}) is stated at each $k$, thus leading to increasing sizes of the information vector. In practice, such a size is limited by looking at a fixed observation horizon of the past, namely: $\Iv_k=[\yv(k-K),...,\yv(k)]$. The size of the information vector, $K$, should be accurately set in order to find a good trade-off between limiting memory usage (which may imply some computational burden for the involved estimation algorithm) and acquiring sufficient knowledge for a reliable decision making.
\subsection{Virtual clock} \label{sec:Virtualclock}
The explicit presence of control variables is disregarded when stating the optimal filter (\ref{eq:OptimalFilter}). The rationale of this choice relies on the fact that the master dynamics is independent of the decision taken at the slaves and slave regulation simply consists of correcting the offset component according to the estimation of $\theta$ and the frequency skew component by using the estimation of $\gamma$ \cite{ToIM}. In this perspective, the concept of \emph{virtual clock} consists of the following: no correction (control variable) is applied to the slave clock register, but a mapping function (i.e., the virtual clock) drives the slave to infer the master notion of time, denoted in the following by $C^v_M$. In other words, a desired actuation may be scheduled at the slave with respect to the last update of $C^v_M$, and a timestamp on an external event may be acquired by a slave and converted in the master timescale using the most recent estimation of $C^v_M$. The virtual clock is updated each time new timestamps can be paired by a slave node. The impact of introducing the control variable in the functional cost, and derive the Riccati equations accordingly, may lead to a smoothing effect on the slave dynamics \cite{Siemens}. Such an effect is not investigated in this work.

\section{Kalman}\label{sec:kalman}
The Kalman filter consists of deriving $\nuv^o(\cdot)$ in closed-form under the mentioned LQG hypotheses. The key issue is to observe that the probability $Pr(\xv(k) | \Iv_k)$ consists of a multidimensional Gaussian function with mean $\muv_k$ and covariance $\Sigmav_k$, $\muv_k$ being the estimation of $\xv(k)$ and $\Sigmav_k$ being the variance of the estimator. Both $\muv_k$ and $\Sigmav_k$ are computed by regular Kalman equations, whose variables depend on the matrices of $\fv(\cdot)$ and $\gv(\cdot)$ and on covariances of $\xiv$ and $\etav$. In this paper, the Kalman equations of \cite{ToIM} are used without reporting them in detail for the sake of synthesis. Very similar equations are used in \cite{Siemens} as well.
\subsection{Remarks}
The Kalman filter guarantees the optimal solution of problem (\ref{eq:OptimalFilter}) (under the LQG hypotheses): at each $k$, the best estimation of $\xv(k)$ is $\muv_k$ and the components of $\Sigmav_k$ decrease to zero with $k \rightarrow \infty$; in other words, the uncertainty on $\xv(k)$, expressed by the difference between $\xv(k)$ and $\muv_k$ under the $\hv(\cdot)$ function, tends to zero with $k \rightarrow \infty$. This holds in stationary conditions, namely, with fixed $\fv(\cdot)$, $\gv(\cdot)$, and stationary noises $\xiv$ and $\etav$. In non-stationary conditions, e.g., if some variance of $\xiv$ or $\etav$ changes, the Kalman filter automatically adapts its calculations and achieves a new optimal solution after a transient period. At each discrete time instant, the Kalman estimation of $\theta$ (denoted by $\hat \theta$) leads to the virtual clock correction, as defined in \ref{sec:Virtualclock}, directly through the application of $\hat \theta$ in (\ref{eq:CSCM}). \\ In passing, it is interesting to note that the Kalman optimality holds for every form of the function $\hv(\cdot)$ in (\ref{eq:OptimalFilter}), since the symmetries of both $\hv(\cdot)$ and $Pr(\xv(k) | \Iv_k)$ (when the latter is Gaussian) let the integration deriving from the $\E(\cdot)$ operator in (\ref{eq:OptimalFilter}) achieve its global minimum exactly when $\muv \rightarrow \xv$. Outside LQG hypotheses, on the other hand, the $Pr(\xv(k) | \Iv_k)$ probability is not Gaussian anymore and its analytical computation, together with its application to the minimization of the above integral, become quite impractical, thus leading to the need of investigating numerical approximation mechanisms.
\subsection{Perfect knowledge of state noises} \label{sec:Perfectknowledge}
The optimality of the Kalman filter is guaranteed with a perfect knowledge of $\xiv$ and $\etav$ covariances. This may be questionable for the $\xiv$ component due to the difficulty of acquiring a perfect knowledge of the XO characteristic curve. While the covariances of the timestamps can be estimated through experiments on the available device, a measurement process of the state noises is a hard task because it involves the direct sampling of the XO physical properties. This is of particular importance with respect to the temperature variations. As the probability distribution of the noise component $\omega_{\gamma^{T}}(k,\cdot)$ is changing in dependence of the environment, one would want to derive an `autonomic' estimation algorithm, capable to adapt to the current conditions (thermal time constant, temperature-frequency XO function (\ref{eq:tempFreqChar}), temperature levels) without manual intervention on the algorithm' parameters in dependence of the environment. This is exactly the final goal of the neural approach.
\subsection{Kalman extensions}
The well-known generalized versions of Kalman filtering, such as the `Extended' and `Unscendent' homonymous filters try to follow non-linear dynamics under the Gaussian hypothesis \cite{ReportCNR}. For this reason, they are disregarded here. Other extensions of the Kalman filters, which have been disregarded for the involved computational burden, but which may be of interest for further performance comparison, are: the on line estimation of the covariances in parallel to the Kalman filter (e.g., \cite{KalmanTuning}) and the Kalman Smoother \cite{KalmanSmoother}, which may help reduce the estimation error through its post-processing of the Kalman filter data.

\section{Regression Splines} \label{sec:splines}
Before addressing directly the numerical approximation of (\ref{eq:OptimalFilter}), simpler heuristics can be formulated. They consist of defining the two-dimensional vector $\varrhov_k=[\hat T_S(k), \hat T_M(k)]$ and a new information vector:
\begin{equation} \label{eq:Is}
\Iv^s_k=[\varrhov_{k-K}, ..., \varrhov_k].
\end{equation}
The $i$-th order spline (denoted by $S_i$) is derived by interpolating, with the $i$-th order, the set of points in $\Iv^s_k$ by means of the Ordinary Least Squares method \cite{OLS}. Points $\varrhov_\kappa$, $\kappa=k-K, ..., k$, lie around the $1-th$ order spline ($S_1$) representing the line of ideal correspondence between master and slave clocks. On the other hand, the distances of $\varrhov_\kappa$ from $S_1$ denotes the quantity of the asynchronism between master and slave. The underlying idea of the splines is therefore to interpolate the trend of those points and to infer the position of the successive points $[\hat T_S(\kappa), \hat T_M(\kappa)]$, $\kappa>k$ as being generated by the spline itself. Being $[\hat T_S(\kappa), \hat T_M(\kappa)]$ representative of $[C_S(\kappa), C_M(\kappa)]$ according to (\ref{eq:measurementnoises1}) and (\ref{eq:measurementnoises2}), the $i$-th order spline becomes a possible virtual clock function ($v_k^{Si}(\cdot)$), as defined in \ref{sec:Virtualclock}, by simply stating: $C^v_M(t)=v_k^{Si}(C_S(t))$, $t>k$. The process of spline interpolation is repeated every $k$: $\Iv^s_k$ moves ahead of one time unit in order to update the calculation of the spline according to the last measurements ($\hat T_S(k+1), \hat T_M(k+1)$) and by disregarding old measurements before $k+1-K$. The continuous updating helps follow a non-stationary behavior of the noises.

In the absence of measurement noise, $\hat T_M(k)=C_M(k)$, $\hat T_S(k)=C_S(k)$ and a perfect estimation is obtained from (\ref{eq:measure1}) and (\ref{eq:measure2}). In the presence of measurement noise, increasing the number of samples $K$ mitigates the effect of oscillations around the spline due to timestamps jitters ($\omega_M$ and $\omega_S$ in (\ref{eq:measurementnoises1}) and (\ref{eq:measurementnoises2})), but, at the same time, the effect of frequency variations due to state noises makes the approximation less accurate. This happens in particular with large $K$ and small orders $i$. The parameter $K$ is thus a compromise between robustness to timestamps jitters and frequency variations.

With sudden frequency variations, spline orders higher than $1$ may perform better because the elements of $\Iv^s_k$ are not linear dependent. The first three order splines are therefore used (linear, quadratic and cubic functions); quadratic and cubic splines try to capture the non-linear behavior of the underlying trend in $\Iv^s_k$. On the other hand, higher order splines have been disregarded to avoid overfitting, to which the splines are more sensitive with more noise and large $K$.

Splines are intrinsically heuristics because they approximate the optimal filter implicitly, without addressing directly the solution of (\ref{eq:OptimalFilter}).

First order regression splines have been successfully used in \cite{2004-FTSP, 2002-RBS, 2012-ETFA-RBIS}; at the best of the authors' knowledge, this is the first time regression splines with order larger than $1$ are used in this context.

An approach similar to linear regression has been applied by \cite{KalmanNTP} to the NTP protocol. Splines surprisingly reveal to be more efficient than expected when the Gaussian hypothesis is not met \cite{KalmanNTP}.

\section{Neural Approximation} \label{sec:neuralApproximation}
A direct way to approximate the optimal filter is now addressed. The proposed method slightly differs from the ones of \cite{MyToNN, IETSensori, ICCPricing, Baglietto}. The following non-linear programming problem is defined:

\begin{eqnarray} \label{eq:NeuralFilter1}
\wv^o=arg \min_{\wv} \E_{\scriptsize {\xv(k)}} \{ \| \xv(k)-\hat \nuv_k(\Iv_k,\wv) \|^2  | \Iv_k \}
\end{eqnarray}
$\hat \nuv_k(\Iv_k,\wv)$ being a neural network with input $\Iv_k$, with output an estimation of $\xv$ and with vector weights $\wv$; its input-output mapping depends also on the number and form of the internal basis function of each layer (e.g., sigmoidal or radial functions).
A first approximation is derived by assuming the optimal filter $\nuv^o(\Iv)$ being obtainable by searching in the space of the functions defined by the chosen structure of $\hat \nuv(\cdot,\wv)$. Another approximation step comes from the observation that the $\E(\cdot)$ operator above and its gradient cannot be obtained in closed-forms. As a result, regular non-linear programming procedures cannot be directly applied to (\ref{eq:NeuralFilter1}). A countermeasure to this consists of resorting to a standard neural network training as follows. The new information vector is defined:

\begin{equation} \label{eq:INN}
\Iv^{NN}_k=[d_{k-K}, ..., d_k]
\end{equation}
$d_k$ being the distance between each $\varrhov_k$ in $\Iv^s_k$ as defined in (\ref{eq:Is}) and the first order spline ($S_1$) interpolating the set of points in $\Iv^s_k$. Then, a training set is stated with $h=1,...,H$ samples of $\Iv^{NN}_k$ as in (\ref{eq:INN}) (being used as input of the neural network in place of $\Iv_k$), and of $e_k= C_M(k)-v_k^{S_1}(k)$, to be used as target of the neural network. Differently from (\ref{eq:NeuralFilter1}), a scalar output is stated for the approximated filter. The problem consists of finding the weights assignment $\wv^o$ so that:
\begin{eqnarray} \label{eq:NeuralFilter2}
\sum_{h=1}^H [^he_k-\hat \nu_k(^h\Iv^{NN}_k,\wv^o)]^2 \leq \rho.
\end{eqnarray}
Problem (\ref{eq:NeuralFilter2}) consists of tuning the output of the neural network (NN) in order to approximate the collected values of $e_k$ as a function of the information vector $\Iv^{NN}_k$. The front end of the neural network thus becomes the synchronization errors introduced by the first order spline. Typical values for the bound $\rho$ are in the range [0.001, 0.5]. After training, the neural network applies the virtual clock as: $C^v_M(t)=v_k^{S_1}(C_S(t))+\hat\nu_k(\Iv^{NN}_k,\wv^o)$, where $k$ is the most recent sample near time $t$. The underlying idea of the method is to follow the concept of approximating the trends in $\Iv^s_k$, as done by the splines, but with a more powerful inference capability than the splines. To this aim, the optimization problem (\ref{eq:NeuralFilter2}) captures the mapping between the actual measurements and the synchronization error. During training, the \textit{NN} should be capable to explore this mapping through the functional dependence on $\Iv^{NN}_k$, which should be more precise than simple interpolation of $\Iv^s_k$.
\remark For the sake of simplicity, the target of the neural estimation focuses on the offset component ($\theta$), without explicitly addressing the skew component ($\gamma$). As evidenced by the following experiments, a significant performance improvement is achieved, despite this simplification. The results also show how the estimation of $\gamma$ may be more crucial to limit the synchronization error between two consecutive synchronization instants with large $\tau$.
\remark The rationale of the adoption of $\Iv^{NN}$ (instead of the one of the optimal filter $\Iv$) relies on the need of deriving neural inputs that do not depend on the absolute value of time. The values collected in $\Iv$ formally depend on the initial instant of time from which the dynamical system is considered; in other words, they depend on the absolute values of the steps $\{0 \cdot \tau, 1 \cdot \tau, 2 \cdot \tau, ... \}$. The values of $\Iv^{NN}$, on the other hand, are relative to the slave time shift through the measured distances from the 1-th order spline.
\remark Both in \cite{MyToNN, IETSensori, Baglietto} and in the present paper, a neural approach is applied to approximate optimal control or state estimation (optimal control in \cite{MyToNN, Baglietto}, state estimation in \cite{IETSensori} and in this paper). The main difference between those works and the idea presented here relies on the numerical approach used to derive the approximation. In \cite{MyToNN, IETSensori, Baglietto}, the functional cost provides indication in the direction of the optimal solution without the explicit knowledge of it. The minimization process is therefore driven by sampling the cost and its gradient and performing a descent step (also known as stochastic gradient \cite{Kushner}). Here, the offline sampling of the optimal solution (i.e., the exact master time) is available. This allows building a training database in which the information vector is mapped onto the difference between master and slave times. A similar approach has been used in \cite{ICCPricing}.
\remark The sequence of approximation steps from (\ref{eq:OptimalFilter}) to (\ref{eq:NeuralFilter2}) leads to the conclusion that evaluating the error computed in approximating the unknown optimal estimation law, $\nuv^o(\cdot)$, with the actual one, $\hat \nu^o(\cdot)$, or, in other words, the convergence of the performance of $\hat \nu^o(\cdot)$ to the one of $\nuv^o(\cdot)$, is a hard task, even though some theoretical results assure the sub-optimal properties of these kinds of approximation schemes. The interested reader is referred to section V of \cite{Baglietto} for an overview or to \cite{Zoppoli} for details on this subject.\\
Despite all the envisaged approximations, the obtained results guarantee a good level of suboptimality and give rise to several insights into the structure of the problem.
\remark Another difficult task arises from finding good generalization capabilities of $\hat \nu(\cdot)$ with respect to non-stationarity conditions, as previously mentioned for the Kalman filter. In this respect, the training is developed with respect to samples coming from non-stationary probability distributions.
This consequently leads to the adaptation of $\hat \nu(\cdot)$ to variable system conditions. An example of this is available in \cite{MyToNN}, in which a similar neural law approximates suboptimal controls in the presence of time-varying conditions. As outlined in subsection \ref{sec:Perfectknowledge}, Kalman does not guarantee this property.

\section{Experiments Setting}\label{sec:experimentsSetting}
\subsection{State parameters}
The stability of a XO highly influences the performance of a CSP. The Allan variance \cite{1990-NIST-clockCharacterization} is the typical parameter used to characterize XOs stability. It depends on the sampling period $\tau$ and can be calculated as $\sigma^2_y(\tau)=\frac{1}{2\tau^2}\left< x_{n-2}-2x_{n+1}+x_n \right>$, where $n$ identifies a specific sample $x_n$ of a temporally ordered sequence of timestamps spaced by an interval $\tau$. The variances ($\sigma^2_{\theta}$ and $\sigma^2_{\gamma}$) of the Gaussian noise components ($\omega_{\theta}$ and $\omega_{\gamma}$) of the clock model of equations (\ref{eq:state1}) and (\ref{eq:state2}) have been set in agreement to the typical values of Allan variance of AT-cut XOs. The first clock model considered models a XO that is not subjected to temperature changes. Other three clock models have been introduced in order to take into account temperature variations.

\begin{itemize}
\item \textit{Gaussian} (`\textit{G}'): $\sigma^2_{\theta}=\unit[10^{-17}]{s^2}$ and $\sigma^2_{\gamma}=\unit[10^{-19}]{}$. The thermal skew component is not preset ($\omega_{\gamma^{T}}(\cdot)=0$). The \textit{G} condition models the instabilities of a XO in a temperature controlled environment. This model takes into account all the typical XOs frequency instabilities (power supply, vibrations, humidity and pressure variations), with the exception of temperature effects. Since $\omega_{\gamma^{T}}(\cdot)=0$, equations (\ref{eq:state1}) and (\ref{eq:state2}) are equal to those reported in \cite{ToIM}. The values of $\sigma^2_{\theta}$ and $\sigma^2_{\gamma}$ are typical of an AT-cut XO of medium stability and they have been calibrated in order to obtain a slightly bigger Allan variance than the one reported in \cite{2011-IJDSN-clockAllan, 2008-ISPCS-clockAllan}.
\item \textit{Temperature} (`\textit{T}'): the frequency skew due to temperature ($\omega_{\gamma^{T}}(\cdot) \neq 0$) is the only effect on quartz stability ($\sigma^2_{\theta}=\unit[0]{}$ and $\sigma^2_{\gamma}=\unit[0]{}$). Although unrealistic, this model is useful to capture the impact of temperature variations without any other source of state noise.
\item \textit{Gaussian+Temperature} (`\textit{G+T}'): it is a realistic condition, in which random noise components of equations (\ref{eq:state1}) and (\ref{eq:state2}) ($\sigma^2_{\theta}=\unit[10^{-17}]{s^2}$ and $\sigma^2_{\gamma}=\unit[10^{-19}]{}$) are taken into account together with the thermal skew component ($\omega_{\gamma^{T}}(\cdot) \neq 0$). The `\textit{G+T}' condition models a XO that is affected by the typical instabilities of a quartz, and it is cyclically exposed to temperature excursions.
\end{itemize}
\begin{table}
  \caption{Simulation parameters and Allan variances for the analyzed clock models}
  \label{tab:simParam}
  \begin{center}
    \begin{tabular}{l|ccccc}
      Scenario & $\sigma^2_{\theta}$ [$\unit[]{s^2}$] & $\sigma^2_{\gamma}$ & $\sigma^2_{\gamma^{T}}(\cdot)$ & $\sigma^2_y(\unit[0.1]{s})$ & $\sigma^2_y(\unit[1.0]{s})$   \\
      \hline\hline
      \textit{G}    & $10^{-17}$ & $10^{-19}$  & $=0$      & $9.97 \cdot 10^{-17}$ & $1.00 \cdot 10^{-17}$  \\
      \textit{T}    & $0$  & $0$              & $\neq 0$  & $6.24 \cdot 10^{-18}$ & $6.37 \cdot 10^{-16}$ \\
      \textit{G+T}  & $10^{-17}$ & $10^{-19}$  & $\neq 0$  & $1.06 \cdot 10^{-16}$ & $6.47 \cdot 10^{-16}$ \\
    \end{tabular}
  \end{center}
\end{table}
\subsection{Other parameters}
\subsubsection{Step size}
The value of the step size $\tau$ represents the sending period of synchronization messages (i.e., common events used by the CSP to evaluate $\hat T_M(k)$ and/or $\hat T_S(k)$), as well as the time step of the clock model of equations (\ref{eq:state1}) and (\ref{eq:state2}).
In practice, synchronization messages are sent periodically every $\unit[0.1- 1]{s}$ and timestamps are frequently acquired in software since devices supporting hardware timestamping are commonly used only for time critical applications, such as the automation of high-voltage substations \cite{substations1}. The use of software timestamp also in this context is becoming an important research challenge \cite{VISS2}. Simulation results are provided for $\tau=\unit[0.1]{s}$ and $\tau=\unit[1]{s}$. Such values are consistent with CSP typical values: $\unit[1]{}$ or $\unit[2]{s}$ for \textit{Sync} messages in IEEE 1588 and for the majority of CSPs and $\unit[100]{ms}$ for RBIS.
The values of Allan variances for the two analyzed $\tau$ and for the three experimental conditions are reported in Table \ref{tab:simParam}.
\subsubsection{Measurement noises}
The measurement jitters ($\omega_M$ and $\omega_S$ in (\ref{eq:measurementnoises1}) and (\ref{eq:measurementnoises2})) are both fixed to a standard deviation of $\unit[1]{\mu s}$, in coherence with the typical accuracy of software timestamps obtained in the \textit{interrupt handler} of the \textit{device driver} of today's real PCs \cite{2008-SIGOPS-interruptLatency}. Experiments performed by the authors on real wireless adapters confirm these results. The maximum uncertainty introduced by the timestamp mechanism could be even in the order of $\unit[10]{\mu s}$ when timestamps are acquired at the communication \textit{socket} level and, in case of high computational load, it can raise up to $\unit[224]{\mu s}$ \cite{PTPeNTP}.
\subsubsection{Neural network}
The neural network used in the experiments (denoted in the pictures with the `\textit{NN}' acronym) has $K$ inputs, $10$ hyperbolic tangent hidden units and one linear output. The input samples $\Iv_k^{NN}$ and the target $e_k$ were normalized by dividing them with the maximum of the absolute value computed over all the input samples and targets of a single train database. The normalization constant computed for each train database is used for the normalization of test samples. It is trained using the standard back-propagation algorithm in $8$ epochs with a \textit{learning rate} linearly decreased during the train from $0.001$ to $0.00001$ and with a \textit{momentum} of $0.01$.
\subsubsection{Kalman}
Differently from the neural network, which is asked to provide estimations over a large set of environments without ad-hoc re-training, the Kalman filter is constantly updated with the right covariances of the state and measurement noises. This is needed each time a new configuration of the temperature noise parameters takes place.
When using Kalman, the first $5000$ samples of each experiment are discarded. This ensures the performance evaluation after the convergence of the filter' parameters. The two approaches are therefore compared unfairly. This helps highlight the robustness of the neural network as any other Kalman filter, which performs on line estimation of the covariances, may hardly guarantee better performance than the one shown here.
\subsubsection{Performance target}
All train and test databases consist of $50000$ samples. The $99.9$ percentile of the synchronization errors is the performance metric. It is denoted by $p99.9$ and represents the $99.9$ percentile of the difference between the real master time $C_M(k)$ and the master time estimated at the slave by using the virtual clock $C^v_M(k)$, in correspondence with the arrival of the synchronization messages. The rationale of this choice relies on the need of capturing the capability of the algorithms to limit master-slave asynchronism below a given threshold, in the 99.9\% of the cases. The qualitative behavior arising from the following results holds for other metrics as well, such as the variance of the synchronization error. The percentile is represented as a function of the size of the information vector $K$. The ranges of $K$ considered in all of the following figures capture the best performance of each algorithm. The experiment related to the `\textit{G+T}' scenario was repeated twice. For the second repetition, denoted as `\textit{G+T+$\tau$}', the $99.9$ percentile is evaluated after $\tau$ seconds the synchronization time instants. The $\tau$ delay represents the maximum time between two synchronization messages (under the hypothesis of an ideal communication network without loss of messages) and the point with higher likelihood of the maximum error. In this respect, `\textit{G+T+$\tau$}' is the worst case scenario.
The estimation of the offset at time $k+\tau$ is obtained assuming a linear behavior of the XO after the last synchronization instant, i.e., by using the last estimations of $\hat\theta(k)$ and $\hat\gamma(k)$ to compute $\hat \theta (k+\tau)= \hat\theta(k)+\hat\gamma(k) \cdot \tau$. For the Kalman filter $\hat\theta(k)$ and $\hat\gamma(k)$ are the outputs of the filter, for the splines $\hat \theta (k+\tau)=v_k^{Si}(k+\tau)$ and for the neural approximation $\hat\gamma(k)=m-1$, where $m$ is the slope of the $1-th$ order spline\footnote{Although the proposed techniques are focused on the estimation of $\theta$, we need for the `\textit{G+T+$\tau$}' scenario the $\gamma$ estimation which is derived from $S_1$ when using the NN. The further extension of the NN for $\gamma$ estimation is an ongoing topic of research.}.

\section{Performance Evaluation}\label{sec:performanceEvaluation}
\begin{figure*}
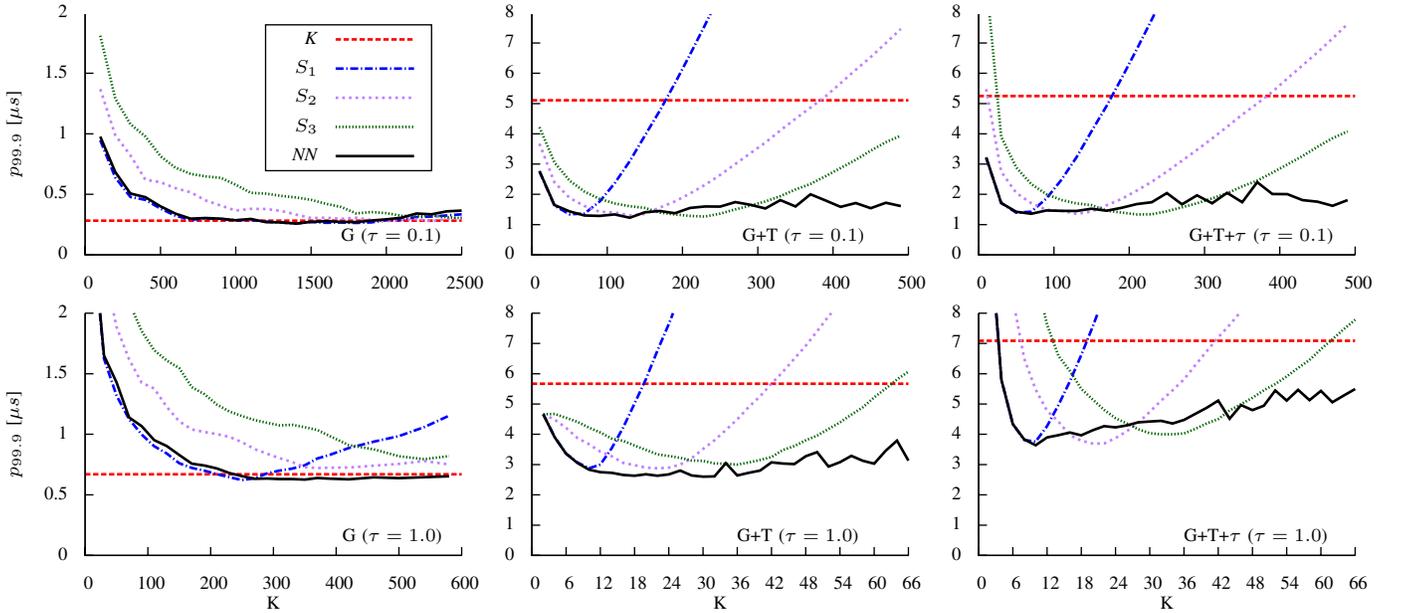

  \scriptsize
  \centering
  \include{fig2}
  \caption{99.9-percentile of the synchronization error, for $K$, $S_1$, $S_2$, $S_3$ and \textit{NN} control algorithms, for the \textit{G}, \textit{G+T} and \textit{G+T+$\tau$} clock models, and with $\tau=\unit[0.1]{s}$ and $\tau=\unit[1.0]{s}$}
  \label{fig:exp1}
\end{figure*}
The results shown here are obtained with the following initial conditions: $\theta(0)\in\unit[[-5, 5]]{s}$ and $\gamma(0)=0$; similar results are obtained with $\gamma(0)=10$ and $\unit[100]{ppm}$. The left column of Fig. \ref{fig:exp1} represents the performance under the first Gaussian (`\textit{G}') scenario. Configuration parameters of the $\omega_{\gamma^{T}}(\cdot)$ function are $tc=\unit[60]{s}$, $p=\unit[600]{s}$, $T_E^{Low}=\unit[-10]{^{\circ}C}$ and $T_E^{High}=\unit[40]{^{\circ}C}$. As expected, Kalman achieves the best performance independently to the setting of $K$ and $\tau$. Such a performance is also reached for $\tau=\unit[0.1]{s}$ by \textit{NN} and $S_1$ for $K>600$ and by $S_2$ and $S_3$ for larger $K$ (1500 and 2000 with $S_2$ and $S_3$, respectively). Similar comments are applicable to the $\tau=\unit[1.0]{s}$ case, a part from $S_1$, which guarantees the lowest percentile only at $K=260$. As mentioned in Section \ref{sec:splines}, splines achieve good performance when $K$ is large enough to obtain a reliable result from the regression applied to the available noisy points and small enough to avoid capturing wrong trends from the data. The $S_1$ with $\tau=\unit[1.0]{s}$ is topical in this perspective because it reveals an increasing sensitivity to noise with $K>260$. The same concept holds true for the middle column of Fig. \ref{fig:exp1} (`\textit{G+T}' scenario), in which all the splines experience a very short range of the optimal $K$. In that case, the \textit{NN} is much more robust to the setting of $K$ and guarantees the best performance because the noise is not Gaussian anymore. The performance gap of Kalman is significant as well. It is also remarkable that the splines' performance is not distant from the best one, in particular with $\tau=\unit[0.1]{s}$. This is an attractive property because splines require much less implementation and computational effort than the neural approach. The synchronization accuracy significantly degrades for all the methods when results are obtained after the synchronization instants, and its worsening is directly proportional with the size of the gap (`\textit{G+T+$\tau$}' scenario). It is more evident with $\tau=\unit[1.0]{s}$, because the estimation of $\hat \gamma$ becomes older. For \textit{NN}, less robustness to $K$ is evidenced for $\tau=\unit[1.0]{s}$ (not for $\tau=\unit[0.1]{s}$). This is due to less accuracy of the $\hat \gamma$ estimation, based on $S_1$, which degrades with increasing $K$. This suggests the extension of the NN including $\hat \gamma$, which can be easily incorporated since the scalar output of the NN should be replaced by a vector addressing both $\hat \theta$ and $\hat \gamma$. The performance of the `\textit{T}' model, mentioned in the previous subsection, is not depicted in Fig. \ref{fig:exp1} because it is very similar to the one of the `\textit{G+T}' case. Overall, the neural approach reveals to be essential to match possible non-Gaussian behaviors of the noises. The splines could be also helpful in this perspective, but they need an accurate tuning of the size of the information vector.

The NN estimation algorithm has been also implemented on an Atmel ATmega328P microcontroller running at $\unit[16]{MHz}$. The aim was to evaluate the inherent computational effort on a microcontroller with the typical complexity of those used in WSN nodes. The results previously obtained with $\tau=\unit[1]{s}$ and $K=15$ have been taken as a reference because they are compatible with the execution of the algorithm in the WSNs context (i.e., with high $\tau$ and small $K$). When setting these parameters on the microcontroller, the registered mean execution time of each iteration involving both features extraction (from $S_1$) and the computation of the output of the trained NN was $\unit[4.932]{ms}$, which is much smaller than the used $\tau$. The algorithm scales linearly with respect to $K$.
\subsection{Generalization capabilities of the \textit{NN}}
\begin{table}
  \caption{The configuration parameters of the function $\omega_{\gamma^{T}}(k,\cdot)$ for the databases used in the simulation.}
  \label{tab:simParameter2}
  \begin{center}
    \begin{tabular}{cc|cccc}
       & Environment & $T^{Low}_E$ $[\unit[]{^{\circ}C}]$ & $T^{High}_E$ $[\unit[]{^{\circ}C}]$ & $tc$ $[\unit[]{s}]$ & $p$ $[\unit[]{s}]$   \\
      \hline\hline
      & \textit{A}  & -10 & 40 & 20 & 200 \\
      Train & \textit{B}  & -10 & 40 & 60 & 600 \\
      & \textit{C}  & -10 & 40 & 180 & 1800 \\
      \hline
      Test & \textit{D} & -5 & 35 & 100 & 1200 \\
      & \textit{E} & -30 & 60 & 10 & 100 \\
      \hline\hline
      & \textit{F} & -30 & 60 & 10 & 100 \\
      Train & \textit{G} & -30 & 60 & 60 & 600 \\
      & \textit{H} & -30 & 60 & 180 & 1800 \\
      \hline
      Test & \textit{I} & -10 & 40  & 20 & 200 \\
           & \textit{L} & -35 & 65  & 20 & 200 \\
      \hline\hline
      \multicolumn{6}{c}{$\sigma^2_{\theta}=\unit[10^{-17}]{(s^2)}$ and $\sigma^2_{\gamma}=\unit[10^{-19}]{(s^2)}$}
    \end{tabular}
  \end{center}
\end{table}
\begin{figure*}
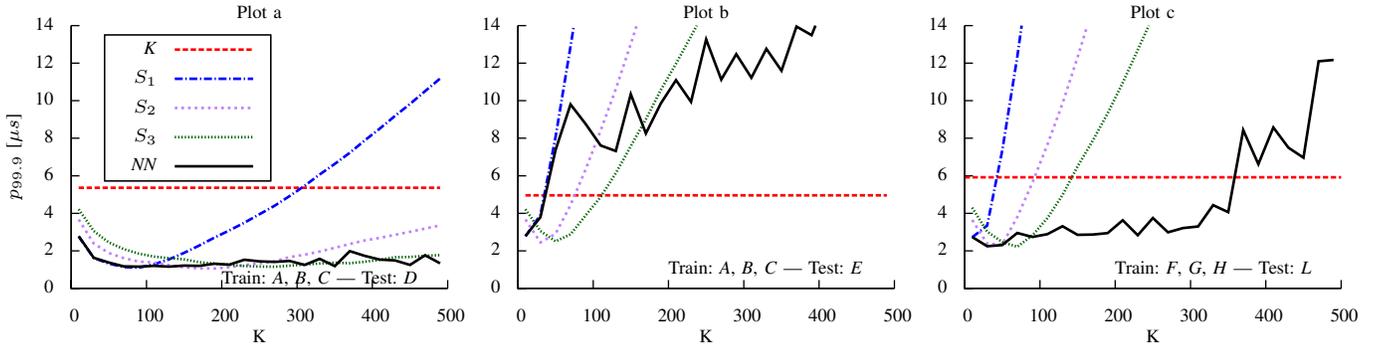

  \scriptsize
  \centering
  \include{fig3}
  \caption{99.9-percentile of the synchronization error for three train and test conditions and for $K$, $S_1$, $S_2$, $S_3$ and \textit{NN} control algorithms ($\tau=\unit[0.1]{s}$).}
  \label{fig:simParameter2}
\end{figure*}
\begin{figure*}
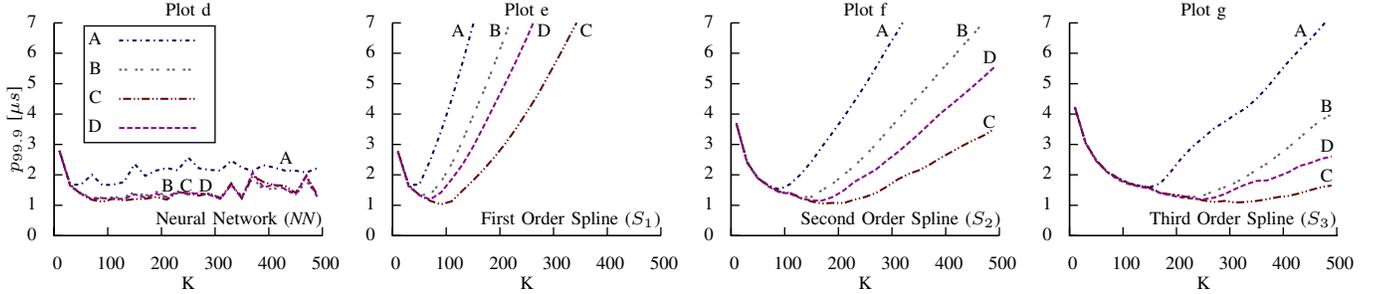

  \scriptsize
  \centering
  \include{fig4}
  \caption{99.9-percentile of the synchronization error for $S_1$, $S_2$, $S_3$ and \textit{NN} control algorithms for conditions \textit{A}, \textit{B}, \textit{C}, \textit{D} ($\tau=\unit[0.1]{s}$).}
  \label{fig:simParameter3}
\end{figure*}
In this set of experiments, the proposed estimation methods are now tested in the `\textit{G+T}' scenario under variable conditions involving temperature thresholds ($T^{High}_E$, $T^{Low}_E$), thermal time constant ($tc$) and object' movements ($p$). Table \ref{tab:simParameter2} summarizes all the tested conditions. The temperature thresholds were chosen to represent environments with medium ($T^{Low}_E=\unit[-10]{^{\circ}C}$ and $T^{High}_E=\unit[40]{^{\circ}C}$) and high temperature excursions ($T^{Low}_E=\unit[-30]{^{\circ}C}$ and $T^{High}_E=\unit[60]{^{\circ}C}$). The time constant, that represents the thermal inertia of the XO, was chosen between two extremes: $tc=\unit[10]{s}$ represents a XO poorly insulated, or placed in a fan cooled case, while $tc=\unit[180]{s}$ is referred to a well thermal insulation. The $p$ value is about ten time $tc$ to allow the XO to reach the new temperature after a temperature change. In such a way, results include all of the temperature excursions between $T^{High}_E$ and $T^{Low}_E$.

In the first plot of Fig. \ref{fig:simParameter2} (plot \textit{a}), the \textit{NN} has been trained with the environmental conditions \textit{A}, \textit{B} and \textit{C}. They represent a range of conditions that should cover all the typical temperature variations to which a XO may be exposed. The test condition \textit{D} represents a situation in which the parameters of the $\omega_{\gamma^{T}}(\cdot)$ noise assume values within the range of the parameters used in the training phase. Despite the parameters of condition \textit{D} are inside the training range, the inherent values denote significantly different environmental conditions involving $T^{High}_E$, $T^{Low}_E$, $p$ and $tc$. Plot \textit{a} helps highlight the generalization capability of the \textit{NN} model. It achieves the lowest $p99.9$, together with the splines, but over a larger range of $K$. It is however worth remembering that only the \textit{NN} method needs a training phase, the splines and Kalman are applied directly during the test phase.
When the test parameters are outside the training range, the \textit{NN} does not generalize well (plot \textit{b}). All the algorithms are however very sensitive to the $K$ parameter in plot \textit{b}. Kalman has a higher $p99.9$ over all the presented settings (plots \textit{a}, \textit{b} and \textit{c}), even if it is always updated with a perfect knowledge of the noises' statistics, which are numerically evaluated in advance.

A way to correct the poor \textit{NN} results of plot \textit{b} is to enlarge the training set even more, by including $T^{High}_E$ and $T^{Low}_E$ of test set \textit{E} and by keeping variable $p$ and $tc$. This corresponds to train conditions \textit{F}, \textit{G} and \textit{H}. Two test sets are used: \textit{I} and \textit{L} (see again Table \ref{tab:simParameter2}).
Test set \textit{I} is equivalent to \textit{D} (i.e., temperature range included in the training set), but with a larger temperature range than \textit{D}.
Test set \textit{L} is equivalent to \textit{E} (i.e., temperature range not included in the training set), but with smaller oscillations outside training. The results of test \textit{I} are satisfactory because they lead to a situation qualitatively very similar to plot \textit{a}; the inherent plot is not reported for the sake of synthesis.
As a results of this, it appears relevant how the \textit{NN} is capable to achieve the best performance over a large range of the $K$ coefficient, together with good generalization capabilities if it operates in environments whose $T^{High}_E$ and $T^{Low}_E$ have been used during training. The results of test \textit{L} are shown in plot \textit{c} of Fig. \ref{fig:simParameter2}.
The values of $T^{High}_E$ and $T^{Low}_E$, which slightly lie outside training, cause performance oscillations of the \textit{NN} with respect to $K$.
Those oscillations do not however lead to a performance decrease as appears in plot \textit{b}. As far as the splines are concerned with respect to tests \textit{I} and \textit{L}, their
minimum reveals again to be very sensitive to $K$ (the \textit{I} case is not reported but it is very similar to the \textit{L} case of plot \textit{c}).

In this perspective, Fig. \ref{fig:simParameter3} highlights the dependence between $p99.9$ and the $K$ coefficient over conditions \textit{A}, \textit{B} and \textit{C}. Similar results are experienced with \textit{F}, \textit{G} and \textit{H}. In the \textit{NN} case, presented in plot \textit{d}, there is a wide range of $K$ values for which the algorithm achieves the minimum of $p99.9$ in all the considered scenarios (\textit{A}, \textit{B} and \textit{C}). The same does not hold true for the splines as evidenced by plots \textit{e}, \textit{f} and \textit{g}. Choosing a unique $K$ for all conditions \textit{A}, \textit{B} and \textit{C} is almost impossible with the splines. For this reason, the \textit{NN} reveals to be the most suitable algorithm for the synchronization problem in the presence of temperature variations. The inherent generalization capabilities guarantee good performance over a large set of environments, this avoiding any on-line adjustment of the algorithm.

\section{Conclusions and Future Work}\label{sec:conclusions}
The paper has presented an innovative estimation approach to clock offset synchronization. It is of interest when the Gaussian hypothesis for clock state equations is not applicable, e.g., in the presence of temperature variations. The approach reveals to be more precise than regular Kalman filtering and more robust to parameters setting with respect to other regression schemes. It reveals to be also applicable to other situations in which other traditional hypotheses, involving linear dynamics or quadratic cost functions, are not met.\\
As the optimal filter is analytically unknown, future work deals with the adoption of other approximating schemes, together with the adoption of other elements of non-linearity, e.g., in other wireless contexts, e.g., \cite{2012-ETFA-RBIS} and by also including some energy metric.

\bibliographystyle{IEEEtran}
\bibliography{IM-13-8286}

\begin{thebibliography}{10}
\providecommand{\url}[1]{#1}
\csname url@samestyle\endcsname
\providecommand{\newblock}{\relax}
\providecommand{\bibinfo}[2]{#2}
\providecommand{\BIBentrySTDinterwordspacing}{\spaceskip=0pt\relax}
\providecommand{\BIBentryALTinterwordstretchfactor}{4}
\providecommand{\BIBentryALTinterwordspacing}{\spaceskip=\fontdimen2\font plus
\BIBentryALTinterwordstretchfactor\fontdimen3\font minus
  \fontdimen4\font\relax}
\providecommand{\BIBforeignlanguage}[2]{{%
\expandafter\ifx\csname l@#1\endcsname\relax
\typeout{** WARNING: IEEEtran.bst: No hyphenation pattern has been}%
\typeout{** loaded for the language `#1'. Using the pattern for}%
\typeout{** the default language instead.}%
\else
\language=\csname l@#1\endcsname
\fi
#2}}
\providecommand{\BIBdecl}{\relax}
\BIBdecl

\bibitem{2013-IEM-CSP1}
G.~Cena, I.~Cibrario~Bertolotti, S.~Scanzio, A.~Valenzano, and C.~Zunino,
  ``{Synchronize your watches: Part I: General-purpose solutions for
  distributed real-time control},'' \emph{IEEE Industrial Electronics
  Magazine}, vol.~7, no.~1, pp. 18--29, 2013.

\bibitem{2013-IEM-CSP2}
------, ``{Synchronize Your Watches: Part II: Special-Purpose Solutions for
  Distributed Real-Time Control},'' \emph{IEEE Industrial Electronics
  Magazine}, vol.~7, no.~2, pp. 27--39, 2013.

\bibitem{2004-TUTORIAL-VIG}
J.~R. Vig, ``{Quartz Crystal Resonators and Oscillators; For Frequency Control
  and Timing Applications - A Tutorial},'' Tutorial, {US Army
  Communications-Electronics Research, Development \& Engineering Center Fort
  Monmouth, NJ, USA, Rev. 8.5.2.2, 2004}.

\bibitem{kalman_bello2}
D.~Fontanelli, D.~Macii, P.~Wolfrum, D.~Obradovic, and G.~Steindl, ``{A clock
  state estimator for PTP time synchronization in harsh environmental
  conditions},'' in \emph{International IEEE Symposium on Precision Clock
  Synchronization for Measurement Control and Communication (ISPCS)}, 2011, pp.
  99--104.

\bibitem{ToIM}
G.~Giorgi and C.~Narduzzi, ``{Performance Analysis of Kalman-Filter-Based Clock
  Synchronization in IEEE 1588 Networks},'' \emph{IEEE Transactions on
  Instrumentation and Measurement}, vol.~60, no.~8, pp. 2902--2909, 2011.

\bibitem{Siemens}
P.~Wolfrum, R.~Scheiterer, and D.~Obradovic, ``{An optimal control approach to
  clock synchronization},'' in \emph{International IEEE Symposium on Precision
  Clock Synchronization for Measurement Control and Communication (ISPCS)},
  2010, pp. 122--128.

\bibitem{kalman_bello}
D.~Fontanelli and D.~Macii, ``{Accurate time synchronization in PTP-based
  industrial networks with long linear paths},'' in \emph{International IEEE
  Symposium on Precision Clock Synchronization for Measurement Control and
  Communication (ISPCS)}, 2010, pp. 97--102.

\bibitem{KalmanNTP}
A.~Bletsas, ``{Evaluation of Kalman filtering for network time keeping},''
  \emph{IEEE Transactions on Ultrasonics, Ferroelectrics and Frequency
  Control}, vol.~52, no.~9, pp. 1452--1460, 2005.

\bibitem{Zoppoli}
R.~Zoppoli, M.~Sanguineti, and T.~Parisini, ``{Approximating networks and
  extended Ritz method for the solution of functional optimization problems},''
  \emph{J. Optim. Theory Appl.}, vol. 112, no.~2, pp. 403--440, Feb. 2002.

\bibitem{RTSPWSNs}
M.~Akhlaq and T.~Sheltami, ``{RTSP: An Accurate and Energy-Efficient Protocol
  for Clock Synchronization in WSNs},'' \emph{IEEE Transactions on
  Instrumentation and Measurement}, vol.~62, no.~3, pp. 578--589, 2013.

\bibitem{2004-FTSP}
M.~Mar\'{o}ti, B.~Kusy, G.~Simon, and A.~L{\'e}deczi, ``{The flooding time
  synchronization protocol},'' in \emph{Proceedings of the 2nd international
  conference on Embedded networked sensor systems (SenSys)}.\hskip 1em plus
  0.5em minus 0.4em\relax New York, NY, USA: ACM, 2004, pp. 39--49.

\bibitem{2002-RBS}
J.~Elson, L.~Girod, and D.~Estrin, ``{Fine-grained network time synchronization
  using reference broadcasts},'' \emph{Oper. Syst. Rev. (SIGOPS)}, vol.~36, pp.
  147--163, Dec. 2002.

\bibitem{2012-ETFA-RBIS}
G.~Cena, S.~Scanzio, A.~Valenzano, and C.~Zunino, ``{The reference-broadcast
  infrastructure synchronization protocol},'' in \emph{IEEE 17th Conference on
  Emerging Technologies Factory Automation (ETFA)}, 2012, pp. 1--4.

\bibitem{SP}
R.~Scheiterer, C.~Na, D.~Obradovic, and G.~Steindl, ``{Synchronization
  Performance of the Precision Time Protocol in Industrial Automation
  Networks},'' \emph{IEEE Transactions on Instrumentation and Measurement},
  vol.~58, no.~6, pp. 1849--1857, 2009.

\bibitem{2008-IEEE-1588std}
IEEE, ``{IEEE Standard for a Precision Clock Synchronization Protocol for
  Networked Measurement and Control Systems},'' \emph{IEEE Std 1588-2008
  (Revision of IEEE Std 1588-2002)}, pp. 1--269, 2008.

\bibitem{1962-IRE-XO}
R.~Bechmann, A.~Ballato, and T.~Lukaszek, ``{Higher-Order Temperature
  Coefficients of the Elastic Stiffinesses and Compliances of Alpha-Quartz},''
  \emph{Proceedings of the IRE}, vol.~50, no.~8, pp. 1812--1822, 1962.

\bibitem{ReportCNR}
\BIBentryALTinterwordspacing
T.~Fiorenzani, C.~Manes, G.~Oriolo, and P.~Peliti, ``{Comparative study of
  unscented kalman filter and extended kalman filter for position/attitude
  estimation in unmanned aerial vehicles},'' \emph{Collana dei Rapporti
  dell'Istituto di Analisi dei Sistemi ed Informatica “Antonio Ruberti”,
  CNR}, 2008. [Online]. Available:
  \url{http://www.iasi.cnr.it/reports/R8008/R8008.html}
\BIBentrySTDinterwordspacing

\bibitem{KalmanTuning}
B.~Akesson, J.~Jorgensen, N.~Poulsen, and S.~Jorgensen, ``{A tool for kalman
  filter tuning},'' \emph{Computer Aided Process Engineering}, vol.~24, pp.
  859--864, 2007.

\bibitem{KalmanSmoother}
A.~Aravkin, ``Robust methods for kalman ﬁltering/smoothing and bundle
  adjustment,'' \emph{PhD Dissertation, Univ. of Washington, Seattle}, 2010.

\bibitem{OLS}
W.~H. Greene, \emph{{Econometric analysis}}.\hskip 1em plus 0.5em minus
  0.4em\relax Prentice Hall, 2002.

\bibitem{MyToNN}
M.~Baglietto, F.~Davoli, M.~Marchese, and M.~Mongelli, ``{Neural approximation
  of open-loop feedback rate control in satellite networks},'' \emph{IEEE
  Transactions on Neural Networks}, vol.~16, no.~5, pp. 1195--1211, 2005.

\bibitem{IETSensori}
F.~Davoli, M.~Marchese, and M.~Mongelli, ``{Non-linear coding and decoding
  strategies exploiting spatial correlation in wireless sensor networks},''
  \emph{IET Communications}, vol.~6, no.~14, pp. 2198--2207, 2012.

\bibitem{ICCPricing}
------, ``{Neural decision making for decentralized pricing-based call
  admission control},'' in \emph{IEEE International Conference on
  Communications (ICC)}, vol.~3, 2005, pp. 1556--1560 Vol. 3.

\bibitem{Baglietto}
M.~Baglietto, T.~Parisini, and R.~Zoppoli, ``{Distributed-information neural
  control: the case of dynamic routing in traffic networks},'' \emph{IEEE
  Transactions on Neural Networks}, vol.~12, no.~3, pp. 485--502, 2001.

\bibitem{Kushner}
H.~Kushner and G.~Yin, \emph{{Stochastic Approximation and Recursive Algorithms
  and Applications}}.\hskip 1em plus 0.5em minus 0.4em\relax Springer, 2003.

\bibitem{1990-NIST-clockCharacterization}
D.~B. Sullivan, D.~W. Allan, D.~A. Howe, and F.~L. Walls, ``{Characterization
  of clocks and Oscillators},'' \emph{NIST, Technical Note 1337}, pp. 1--342,
  1990.

\bibitem{2011-IJDSN-clockAllan}
G.~Gaderer, A.~Nagy, and P.~Loschmidt, ``{Achieving a Realistic Notion of Time
  in Discrete Event Simulation},'' \emph{International Journal of Distributed
  Sensor Networks}, vol. 2011, pp. 1--11, 2011.

\bibitem{2008-ISPCS-clockAllan}
P.~Loschmidt, R.~Exel, A.~Nagy, and G.~Gaderer, ``{Limits of synchronization
  accuracy using hardware support in IEEE 1588},'' in \emph{IEEE International
  Symposium on Precision Clock Synchronization for Measurement, Control and
  Communication (ISPCS)}, 2008, pp. 12--16.

\bibitem{substations1}
D.~Ingram, P.~Schaub, and D.~Campbell, ``{Use of Precision Time Protocol to
  Synchronize Sampled-Value Process Buses},'' \emph{IEEE Transactions on
  Instrumentation and Measurement}, vol.~61, no.~5, pp. 1173--1180, 2012.

\bibitem{VISS2}
M.~Lixia, A.~Benigni, A.~Flammini, C.~Muscas, F.~Ponci, and A.~Monti, ``{A
  Software-Only PTP Synchronization for Power System State Estimation With
  PMUs},'' \emph{IEEE Transactions on Instrumentation and Measurement},
  vol.~61, no.~5, pp. 1476--1485, 2012.

\bibitem{2008-SIGOPS-interruptLatency}
P.~Regnier, G.~Lima, and L.~Barreto, ``{Evaluation of interrupt handling
  timeliness in real-time Linux operating systems},'' \emph{Oper. Syst. Rev.
  (SIGOPS)}, vol.~42, no.~6, pp. 52--63, Oct. 2008.

\bibitem{PTPeNTP}
P.~Ferrari, A.~Flammini, S.~Rinaldi, A.~Bondavalli, and F.~Brancati,
  ``{Experimental Characterization of Uncertainty Sources in a Software-Only
  Synchronization System},'' \emph{IEEE Transactions on Instrumentation and
  Measurement}, vol.~61, no.~5, pp. 1512--1521, 2012.

\end{thebibliography}

\end{document}